\documentclass[%
 reprint,
superscriptaddress,
 amsmath,amssymb,
 aps,
]{revtex4-2}

\usepackage{graphicx}
\usepackage{dcolumn}
\usepackage{bm}
\usepackage{subcaption}
\usepackage{stackengine}
\usepackage{microtype}
\usepackage[utf8]{inputenc}
\usepackage[breaklinks=true]{hyperref}

\usepackage{xcolor}
\hypersetup{
    colorlinks,
    linkcolor={red!50!black},
    citecolor={blue!50!black},
    urlcolor={blue!80!black}
}

\begin{document}

\title{Effect of thresholding on avalanches and their clustering \\ for interfaces with long-range elasticity}

\author{Juha Savolainen}
\affiliation{Aalto University, Department of Applied Physics, PO Box 11000, 00076 Aalto, Finland}
\author{Lasse Laurson}
\affiliation{Computational Physics Laboratory, Tampere University, P.O. Box 692, FI-33101 Tampere, Finland}
\author{Mikko Alava}
\affiliation{Aalto University, Department of Applied Physics, PO Box 11000, 00076 Aalto, Finland}
\affiliation{NOMATEN Centre of Excellence, National Centre for Nuclear Research, A. Soltana 7, 05-400 Otwock-Swierk, Poland}

\date{\today}

\begin{abstract}

Avalanches are often defined as signals higher than some detection level in bursty systems. The choice of the detection threshold affects the number of avalanches, but it can also affect their temporal correlations. We simulated the depinning of a long-range elastic interface and applied different thresholds including a zero one on the data to see how the sizes and durations of events change and how this affects temporal avalanche clustering. Higher thresholds result in steeper size and duration distributions and cause the avalanches to cluster temporally. Using methods from seismology, the frequency of the events in the clusters was found to decrease as a power-law of time, and the size of an event in a cluster was found to help predict how many events it is followed by. The results bring closer theoretical studies of this class of models to real experiments, but also highlight how different phenomena can be obtained from the same set of data.

\end{abstract}

\maketitle

\section{Introduction}

\begin{figure}
\includegraphics[width=\linewidth]{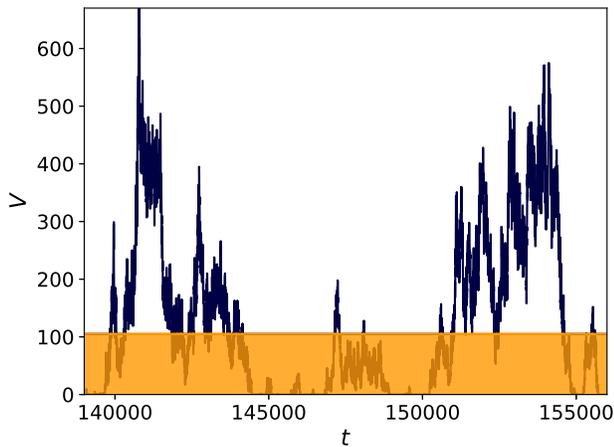}
\caption{\label{fig:threshold} A snapshot of simulated avalanches with a visualized detection threshold. The dark blue line shows the velocity $V$ of the interface as a function of the simulation time $t$, and the orange region depicts the threshold. When a threshold is used, only movement above it is considered, so whenever the velocity signal goes inside the orange region, the velocity is set to zero.}
\end{figure}

A slide in a sandpile \cite{sandpile}, Barkhausen noise in magnets \cite{Durin2005TheBE}, and solar flares \cite{Aschwanden_2014} are examples of avalanches in physics. Avalanches are intermittent events with scale-free sizes and durations, defined as the events that are large enough to stand out from some background activity or noise. The choice of which events are large enough is done by setting a detection threshold, below which all data are ignored.

Filtering out small signals affects also the larger events, as has been shown for random walks \cite{Laurson_2009, Font_Clos_2015, time-seriesthresholding} and elastic interfaces \cite{Jani_evi__2016}. In experiments however a threshold might be unavoidable, as even if all background activity could be removed from the data by other means, the detection devices might not be able to record the smallest relevant signals. Therefore, it is both interesting and useful to study how a threshold affects the results in different systems.

Elastic interfaces model for example magnetic domain walls \cite{DomainWallMagneticFilm, Dynamicsferromagneticdomainwall} and fluid invasion in porous media \cite{Fluidinvasion, Imbibitionfronts}. We use a similar system to what was used in \cite{Jani_evi__2016}, in which the elastic interactions are long-ranged with a quadratic decay. For example planar cracks in fracture mechanics  \cite{Rice}, contact lines in wetting \cite{ContactAngle}, and low-angle grain boundaries in dislocations \cite{Low-angleGrainB} exhibit this type of elasticity.

Planar cracks, as the name suggests, are tears that move in a plane in a material. In experiments they can be created by pulling apart an object with a pre-existing crack. The front of the crack behaves like an elastic interface that moves intermittently whenever the pulling force is enough to overcome a weak spot in the object. As one part of the crack front moves forward, it tends to pull neighbouring parts with it, creating avalanches in the movement. \cite{GaoRice, CriticalTestoftheTheory} The crack front of course deviates from perfectly planar movement, but the phenomenon was also demonstrated by attaching two sandblasted Plexiglas plates on top of each other and tearing them apart \cite{plexi, PhysRevE.83.046108}.

Tools from seismology are often borrowed to study correlations in avalanches \cite{spatiotemporal, creepmotionof, M_kinen2020}. The fracture mechanics model \cite{Bar_s_2018, Bar_s_2018b, Bar_s_2019}, as well as other phenomena like wood compression \cite{M_kinen_2015}, follow similar scaling laws as what are found for earthquakes.

In seismology, earthquakes divide into so called mainshocks and aftershocks. A mainshock is an event that triggers smaller earthquakes in the nearby region, and the aftershocks are the triggered events. The productivity law states that the number of triggered aftershocks grows exponentially with the magnitude of the mainshock, or equivalently as a power law of the mainshock's energy. The Omori-Utsu law states that the frequency of the aftershocks decreases as a power law of the time elapsed after the mainshock. There are also small events known as foreshocks that precede mainshocks. \cite{Utsu1970, Utsu1995TheCO}

Barés et al. used a similar division into mainshocks and aftershocks for activity in the interface model and the related planar crack experiment, treating event sizes analogously to the energies of earthquakes \cite{Bar_s_2018, Bar_s_2018b, Bar_s_2019}. The avalanches followed the productivity law, the Omori-Utsu law, and a law called Båth's law. Båth's law states that the magnitude of a mainshock is on average 1.2 times the magnitude of its largest aftershock, regardless of the mainshock's magnitude.

The most obvious side effect of a threshold is that it makes avalanches smaller by removing a part of the movement. The smallest avalanches vanish completely, which reduces the number of events. A perhaps more interesting effect is that different peaks of the same event can get labelled as separate events, as every time the velocity drops below the detection threshold and comes back up, the avalanche is assumed to have stopped and a new one to have initiated. Thus, a threshold both removes events and creates new ones. 

A threshold creates power law distributed waiting times between avalanches in the interface model \cite{Jani_evi__2016}. We expect a threshold to also affect the analogues of the productivity and Omori-Utsu laws, as the choice of a threshold affects how many events and thus aftershocks arise from an underlying signal.

In subsection \ref{results1} we look at how a threshold affects the size and duration distributions of avalanches, as well as repeat the earlier results found for the waiting times in \cite{Jani_evi__2016}. Subsection \ref{Omorisubsection} discusses the frequency of avalanches and aftershocks. In subsection \ref{prodsubsection} the productivity law is looked at with two different definitions for the aftershocks. First, the aftershocks are defined similarly as in \cite{Bar_s_2018, Bar_s_2018b, Bar_s_2019}, and then the aftershocks are required to be within a specific window of time after the mainshock.

\section{The numerical model}

We simulated the movement of a long range elastic interface around the depinning point, which is the point where the system is driven just enough to cause movement, with a cellular automaton model. The interface consists of $L=2^{17}$ points moving in a direction perpendicular to the initial direction of the interface. 
Each point experiences the same external driving force and individual pinning and elastic forces. The pinning force for each point is a Gaussian random variable with variance 0.3, and it changes every time the point moves. 

The elastic force depends on how far each point has advanced, and it uses the quadratically decaying form
\begin{equation}
    \label{eqn:Rice}
    f_i=k\sum_{j\neq i}\dfrac{h_j-h_i}{(j-i)^2},
\end{equation}
where $k=0.3$ is a spring constant and $h_l$ denotes how many steps has the point at site $l$ moved. The sum is over all points of the interface. Periodic boundary conditions modify the elastic term to 
\begin{equation}
    \label{eqn:periodic}
    f_i = \dfrac{\pi^2 k}{L^2}\sum_{j \neq i}\dfrac {h_j -h_i} {\sin^2 \Big( \dfrac{\pi}{L}(j-i) \Big)}.
\end{equation}
using $\sin^{-2}x=\sum_{n={-\infty}}^{\infty}(n\pi+x)^{-2}$.

Each time step starts by calculating the elastic force for all the points. Each point for which the sum of the elastic, pinning, and driving forces is positive moves one step.

The interface starts with a straight configuration, so it experiences a large initial movement until the elastic forces grow large enough to balance out the pinning forces. This initial roughening is not included in the data. When the system stops after the initial reconfiguration, the external driving is increased until at least one point becomes unstable and the first recorded avalanche initiates.

The implementation for the external force is somewhat simpler than the common comoving approach, in which the interface follows an average velocity set by the experiment or simulation with a set spring tension \cite{PhysRevLett.101.045501, CreepandDep}. Now the driving force changes with a constant rate at each time step, so that during timesteps when at least one point in the interface moves, the driving decreases by $10^{-7}$, and at quiescent steps the driving increases by $10^{-7}$. This way the driving force balances as close to a theoretical critical value as possible, and as a result roughly half of the timesteps contain movement. The naturally occurring waiting times between events allow us to study avalanches with no threshold at all.

The simulations run for $2^{18}$ timesteps. The data are averaged over 100 runs. The size of an avalanche is how much the sum of all $h_l$ changed, i.e., how much the interface moved in total. A threshold subtracts a constant number of movement from each timestep as long as the result is not negative. Durations and waiting times are the number of time steps spent above or below the threshold in simulation time.

\section{Results}

\begin{figure}
\includegraphics[width=\linewidth]{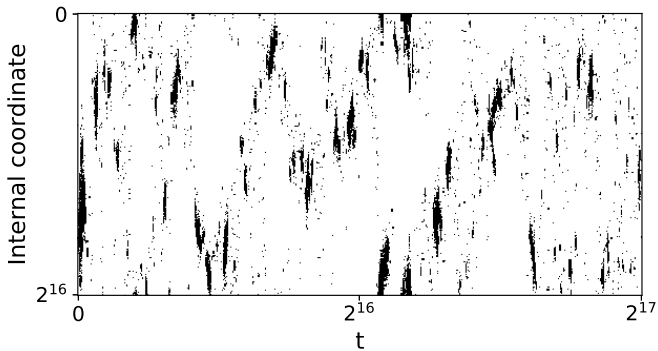}
\caption{\label{fig:Map} A space-time map of the avalanches during one simulation. The horizontal axis is simulation time and the vertical axis is position along the interface. The black dots denote which parts of the interface moved at that time. Note that there are periodic boundary conditions, so the points in the upper and lower boundaries in the graph are next to each other.}
\end{figure}

\begin{figure}
\includegraphics[width=\linewidth]{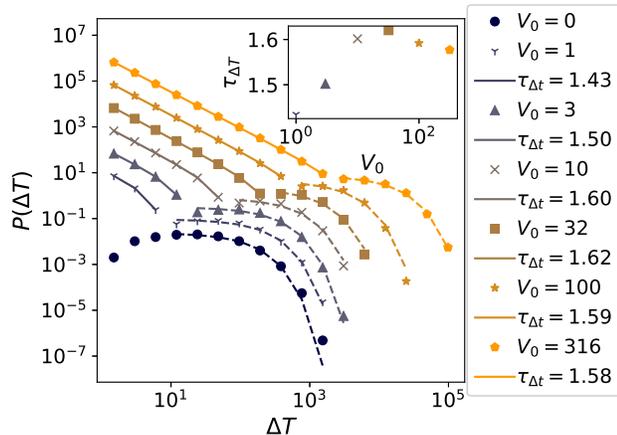}
\caption{\label{fig:waitingt} The waiting time distribution. The continuous lines are power-law fits with functions $\propto (\Delta t)^{-\tau_{\Delta t}}$ and the dashed lines are exponent function fits. The lowermost graph shows that without a threshold the waiting times follow an exponential distribution, and moving upwards the graphs start exhibiting a power-law region which grows with the threshold. Note that the power-law region starts forming already in the second graph with threshold $V_0=1$, although the fitted exponent is very inaccurate due to the limited number of datapoints. The power-law distribution describes the waiting times between the sub-events created with the threshold. The original events get further away from each other as the threshold increases, and consequently the exponentially distributed region moves to longer times. As before, the graphs have been moved vertically for visual clarity. The legend and the inset show the thresholds and the fitted exponents for the power-law region.}
\end{figure}

Each dataset contains 1174.53 avalanches on average. The average signal is 52.9 and during avalanches the average signal is 104.5. The main results are in Figures \ref{fig:sizes}-\ref{fig:KaikkiProd}, which show different avalanche distributions using thresholds 0, 1, 3, 10, 32, 100, and 316.

Figure \ref{fig:Map} shows the spatial and temporal distribution of the activity in one simulation. All simultaneous movement belongs to the same avalanche, even if there is a large spatial separation, as even distant points have direct elastic interactions. The avalanches consist of clusters of movement that are dense in the middle and turn into sparse clouds farther away. Adding a threshold to the global movement signal might have a similar effect as removing some movement of the remote points. The remote points cause avalanches to start and end more smoothly, and possibly unify the dense cores of avalanches that are not simultaneous.

\subsection{\label{results1}Increased number of small events}

\begin{figure}
\includegraphics[width=\linewidth]{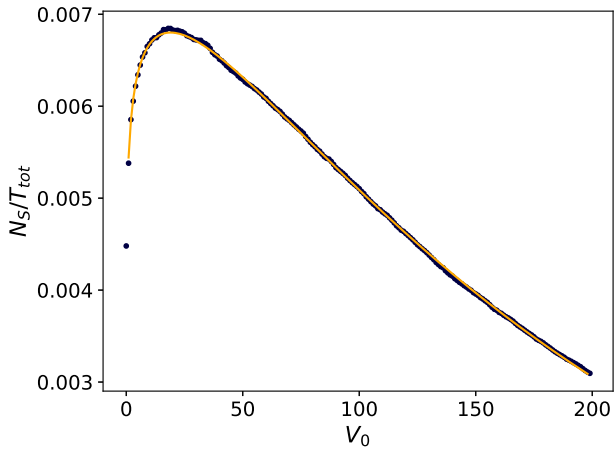}
\caption{\label{fig:Nevents} The number of events $N_S$ per the simulation's duration $T_{tot}$ as a function of the threshold $V_0$. The number is the highest at threshold 18. The continuous line is a fit by a function $\propto V_0^A e^{-BV_0}$, where $A \approx 0.11$ and $B \approx 0.0059$ are constants.}
\end{figure}

\begin{figure}
\includegraphics[width=\linewidth]{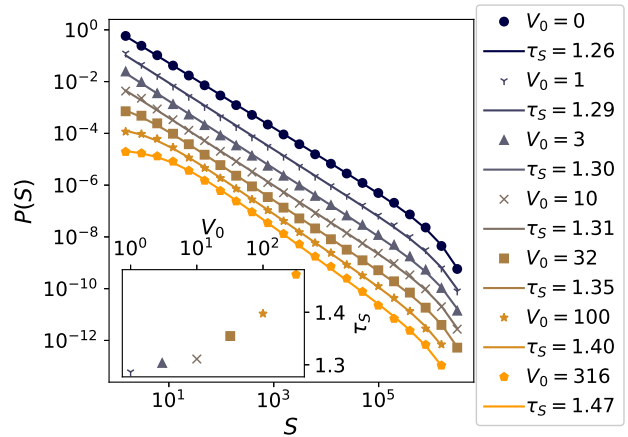}
\caption{\label{fig:sizes} The size distribution of the avalanches fitted as $\propto (1+S/S_{min})^{-\tau_S} e^{-S/S_{max}}$, where $S$ is size, $\tau_S$, is the power law exponent, and $S_{min}$ and $S_{max}$ are the cutoffs at small and large avalanches. The different graphs represent different thresholds. The graphs have been shifted vertically to avoid overlapping. The legend and the inset show the thresholds and the fitted exponents.}
\end{figure}

\begin{figure}
\includegraphics[width=\linewidth]{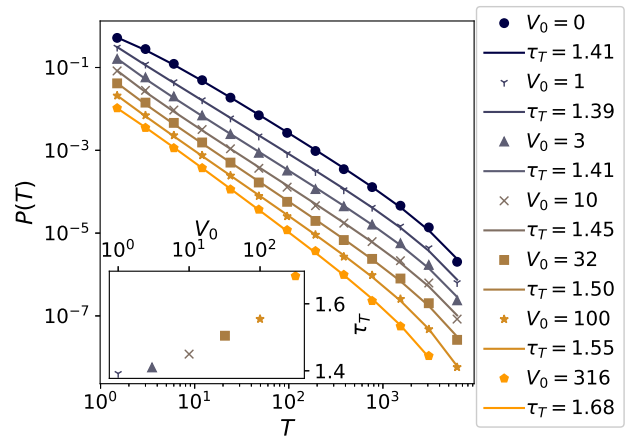}
\caption{\label{fig:durations} The duration distribution fitted using a similar function $\propto (1+T/T_{min})^{-\tau_T} e^{-T/T{max}}$ as for the size distribution. Again, the different graphs represent different thresholds and the legend and the inset show the thresholds and the fitted exponents. The graphs have been moved vertically to avoid overlapping.}
\end{figure}

Figure \ref{fig:Nevents} shows the number of events at each threshold. At small thresholds the number is growing, until it starts decreasing exponentially after its peak when the threshold equals 18. The number stays above its original value until the threshold is increased to 126.

The change in the number of events has an effect on the size and duration distributions, shown in Figures \ref{fig:sizes} and \ref{fig:durations} respectively. As the size of every event decreases and small events are both destroyed and created, the net effect is an increase in small and short events and a decrease in the larger and longer ones. Thus the magnitudes of the size and duration distributions' power-law exponents increase with the threshold. At thresholds close to the average velocity of 105 during avalanches, the exponents of the size and duration distributions change by roughly 10 percentages compared to the zero-threshold graphs. Consequently, experiments should yield slightly larger exponents than what are found in theoretical studies that do not necessarily require a threshold.

Similarly as in \cite{Jani_evi__2016}, the waiting time distribution changes from an exponential one into a power law with an exponential bump at the end. As shown in \cite{dbbf0c72c2ec46f88c9edae0aaad21e5} and \cite{Dobrinevski_2014}, avalanches start and end, on average, with slower movement. Thus a threshold typically cuts out the beginning and the end of the events, increasing the waiting times between the original avalanches. Because of this, the exponential waiting time region starts at later times as the threshold increases. The new events created by splitting the original avalanches on the other hand must have waiting times shorter than the avalanche durations, so they fill the short time-scales in the waiting time distribution. 

Interestingly, the power-law region starts forming already at threshold $V_0=1$, which is the smallest non-zero velocity that the interface can have. Therefore, any choice of a threshold in an experiment should create an increase in the waiting time distribution for at least the smallest values.

As the threshold increases, the amount of datapoints in the power-law regions in the waiting time distributions grows, and the exponents in the duration and waiting time distributions both approach $1.6$. This means that the interface velocity makes symmetric visits above and below a large threshold before the underlying event ends. In other words, at large velocities the velocity starts to resemble a symmetric random walk, as discussed in \cite{Jani_evi__2016}.

\subsection{\label{Omorisubsection}Temporal clustering of events}

\def\stackalignment{l}

\begin{figure*}

  \begin{subfigure}{.49\linewidth}
    \topinset{(a)}{\centering\includegraphics[width=\linewidth]{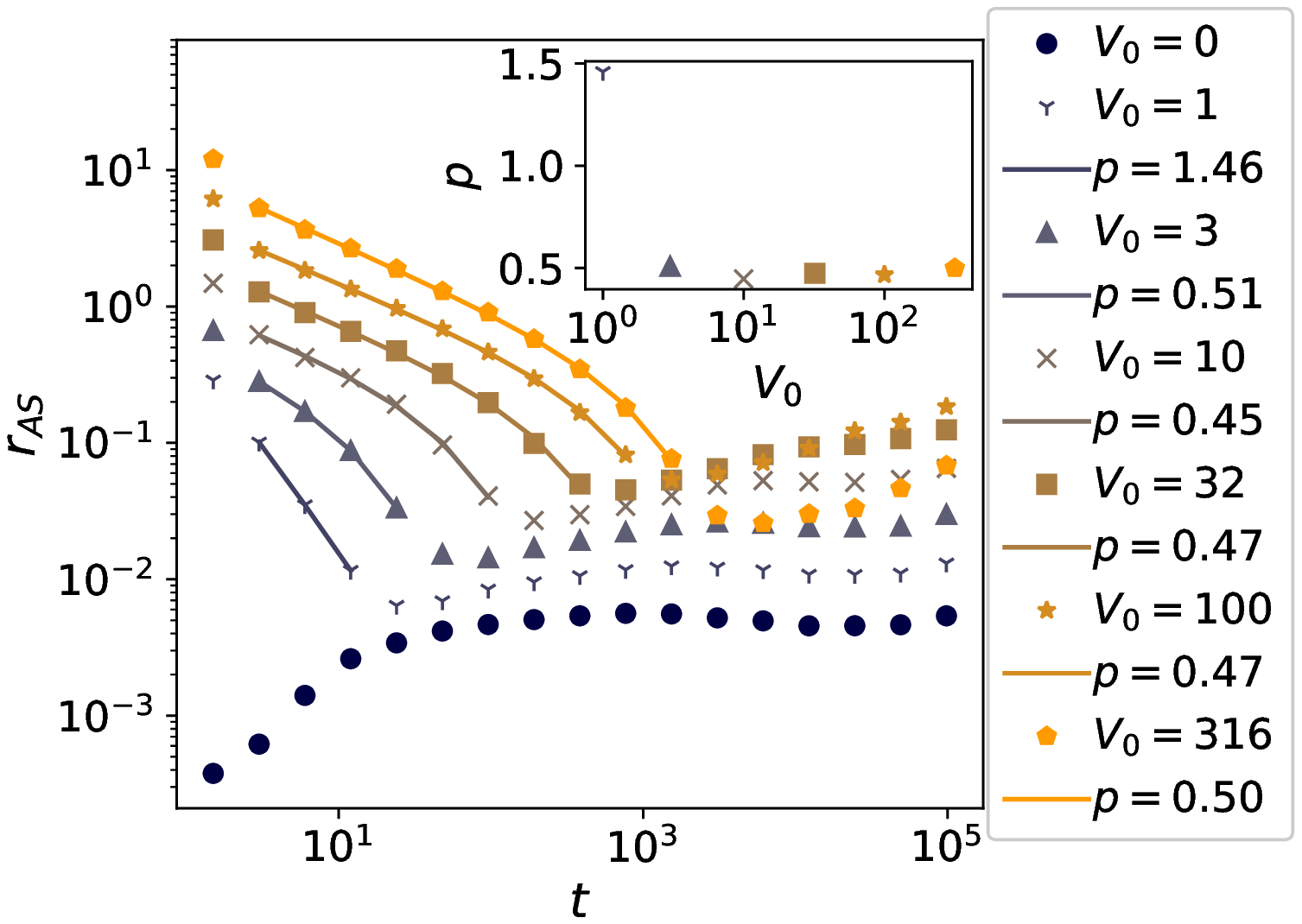}}{0.3in}{0in}
    \phantomcaption{\label{fig:Omori}}
  \end{subfigure}
  \begin{subfigure}{.49\linewidth}
    \topinset{(b)}{\centering\includegraphics[width=\linewidth]{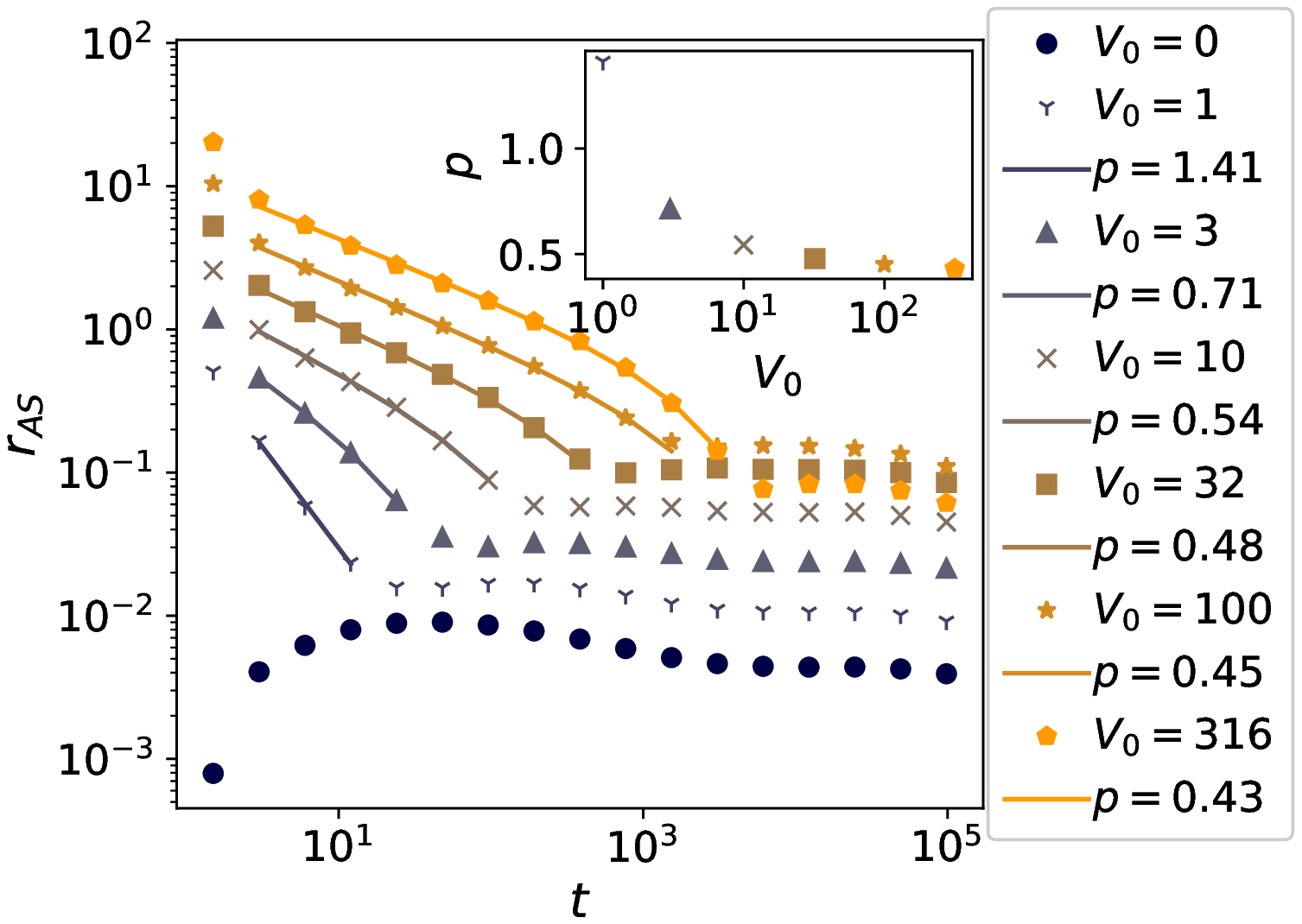}}{0.3in}{0in}
    \phantomcaption{\label{fig:Omori2}}
  \end{subfigure}
  
  \caption{\label{fig:Omorit}The frequency of aftershocks and clustered avalanches as a function of time. Figure \ref{fig:Omori} shows the rate of aftershocks after a mainshock, and Figure \ref{fig:Omori2} shows the rate of avalanches after any avalanche. The continuous lines are fits using a function $\propto t^{-p} e^{-t/t_P}$, where $t$ is time and $p$ and $t_P$ are constants. As previously, the different graphs show different thresholds, and they have been shifted vertically for clarity. The legend and the inset show the thresholds and the fitted exponents. However the fits for small thresholds are very inaccurate due to the small number of fitted data points.} 

\end{figure*}

The division of avalanches into series of smaller ones changes the temporal clustering of events. Barés et al. studied the clustering of avalanches in elastic interfaces with the concept of mainshocks and aftershocks used in seismology \cite{Bar_s_2019, Bar_s_2018, Bar_s_2018b}. Any event could take the role of a mainshock, and after that all the subsequent events were labelled as aftershocks, until an event at least as large as the mainshock was encountered. Seismologists often require the aftershocks to be within some distance of the mainshock \cite{10.1093/gji/ggaa252, isearthquaketriggeringdrivenby}, but that is not feasible in the interface problem, when only the velocity of the whole interface is looked at, and not local movement. 

The productivity law in seismology means that the number of aftershocks that follow a mainshock is proportional to a power of the mainshock's energy. The Omori-Utsu law states that frequency of the aftershocks decreases as a power of the time elapsed after the triggering event. \cite{Utsu1970, Utsu1995TheCO} Barés et al. found that similar laws also applied to the mainshocks and aftershocks in interface dynamics. The number of aftershocks was proportional to a power of the mainshocks' size, and the aftershock frequency decreased as a power of time.

Figure \ref{fig:Omori} shows the aftershock frequency in our system, with the definition that all shocks after a mainshock are aftershocks, until a shock at least as large as the mainshock is encountered. Interestingly, we find a decreasing aftershock frequency only when using a threshold. As the waiting times in the underlying pure signal showed no correlations, the frequency of the events without a threshold only increases with time, possibly as more exponentially distributed waiting times have ended and new avalanches initiated.

Just as for the waiting times, already the minimal positive threshold $V_0=1$ causes a dramatic increase in the aftershock frequency for small times. With higher thresholds, the increased activity extends to longer times, and a power-law region starts forming.

Contrary to the findings of Barés et al., we see a plateau and even a slight increase in the aftershock frequencies for longer times. As the increased activity results from a threshold dividing underlying avalanches, the rate of events initially decreases as more of the avalanches in the pure signal have ended. Then as the waiting times in the underlying signal end and new avalanches begin, the aftershock frequency for a thresholded signal plateaus and possibly grows if there are enough new avalanches to divide.

Since the increased frequency of events seems to arise from the altered waiting time distribution, we should get similar results even without dividing the avalanches into mainshocks and aftershocks. Figure \ref{fig:Omori2} shows the average rate of events after each event, without requiring the following events to be smaller than the initial one.

The event frequency looks very similar to Figure \ref{fig:Omori} with limited sized aftershocks. However, the increased amount of data delay the cutoffs in the graphs, making the power-law fits more reasonable and also altering the exponents. Now the fitted exponents decrease monotonously with the threshold, approaching $0.4$ for the largest thresholds.

Similarly to the durations of the avalanches in the underlying pure signal, the durations of the avalanche clusters in the thresholded data probably also follow a decreasing distribution. As the number of active avalanche clusters decreases, the average frequency of avalanches decreases, causing the decreasing rate of events in Figures \ref{fig:Omori} and \ref{fig:Omori2}.

\subsection{\label{prodsubsection}Number of aftershocks}

\begin{figure*}

  \begin{subfigure}{.49\linewidth}
    \topinset{(a)}{\centering\includegraphics[width=\linewidth]{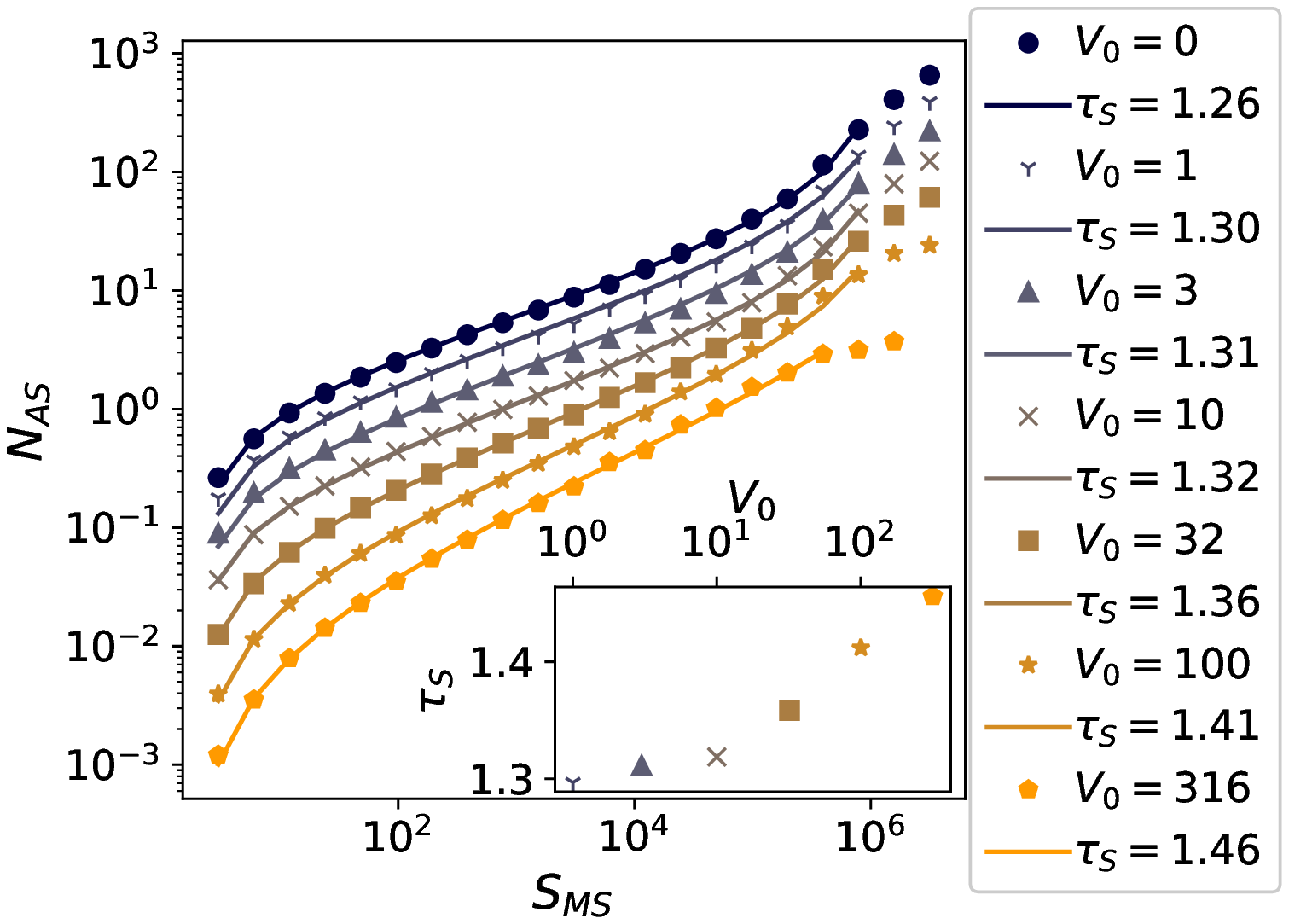}}{0.3in}{0in}
    \phantomcaption{\label{fig:NAS}}
  \end{subfigure}
  \begin{subfigure}{.49\linewidth}
    \topinset{(b)}{\centering\includegraphics[width=\linewidth]{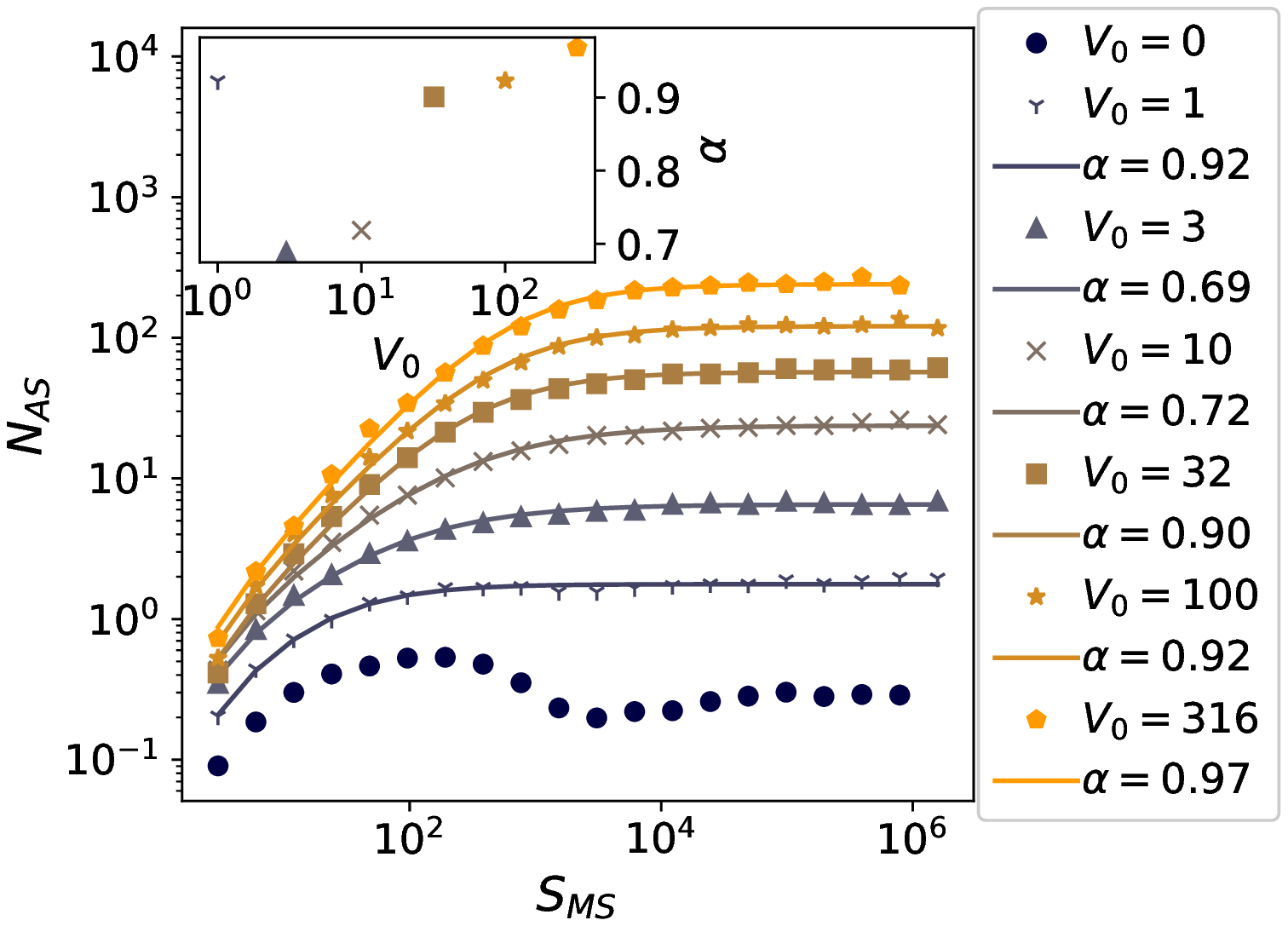}}{0.3in}{0in}
    \phantomcaption{\label{fig:NAS100}}
  \end{subfigure}
  
\vspace{-1.5\baselineskip}

  \begin{subfigure}{.49\linewidth}
    \topinset{(c)}{\centering\includegraphics[width=\linewidth]{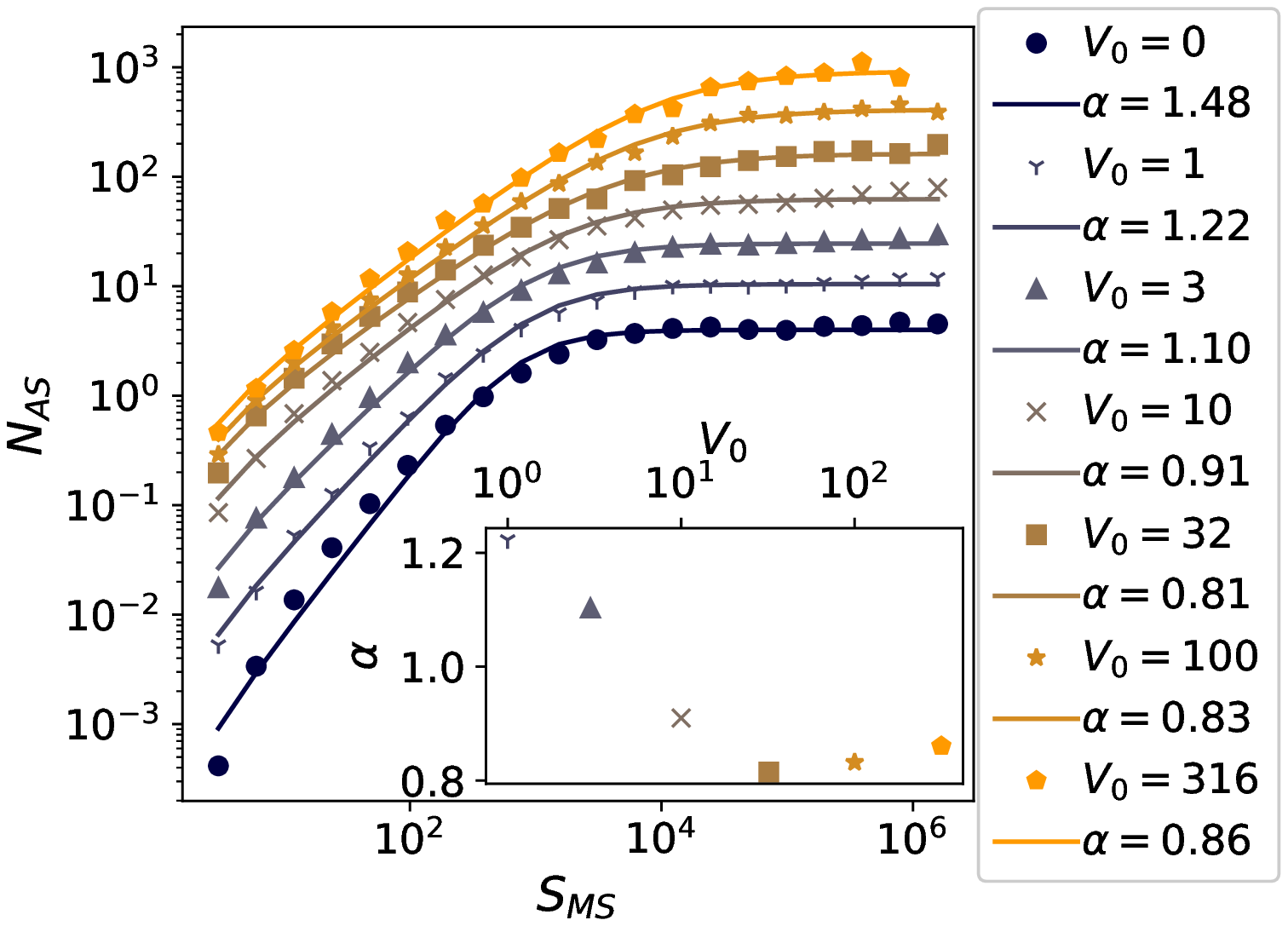}}{0.3in}{0in}
    \phantomcaption{\label{fig:NAS1000}}
  \end{subfigure}
  \begin{subfigure}{.49\linewidth}
    \topinset{(d)}{\centering\includegraphics[width=\linewidth]{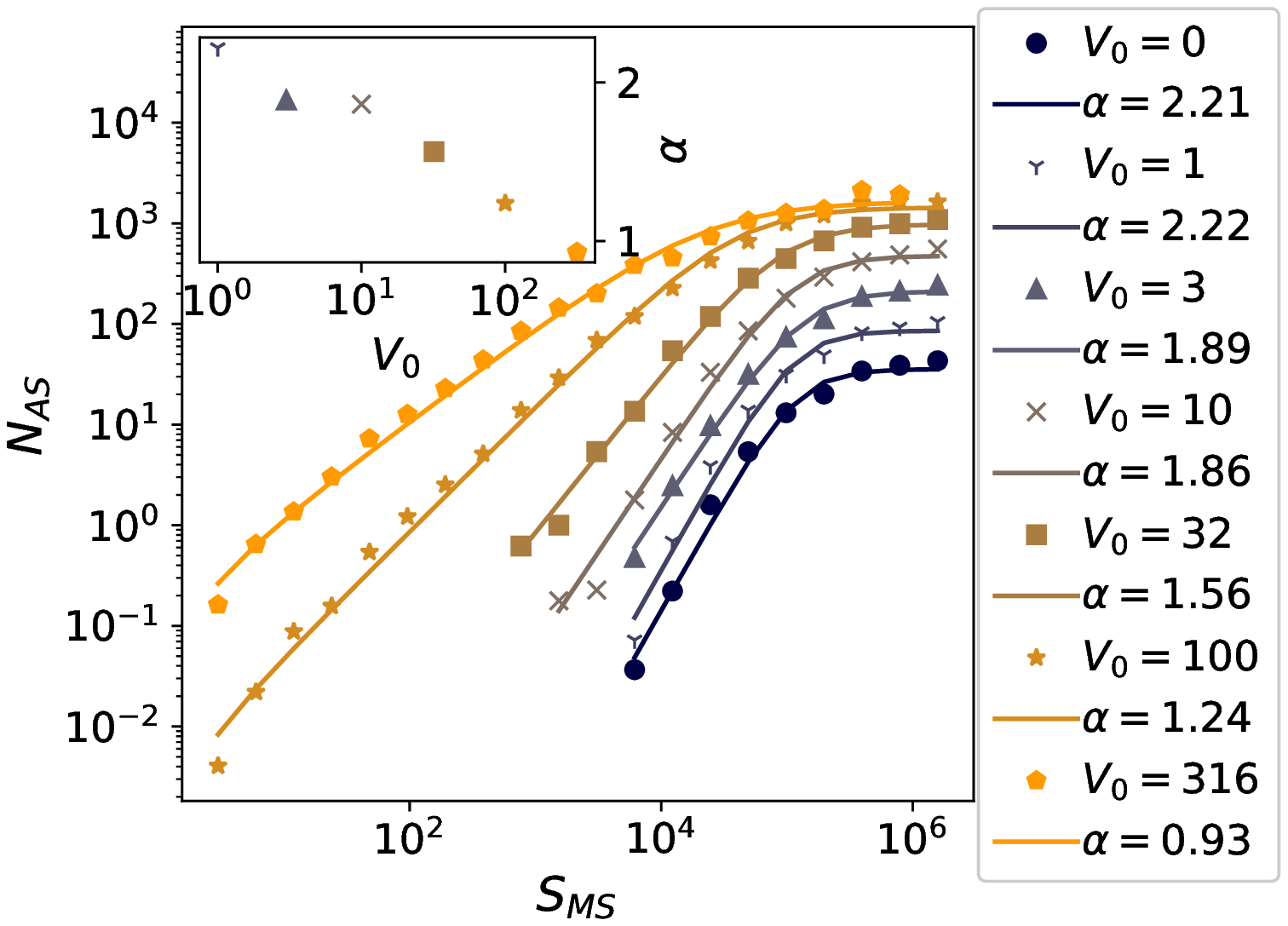}}{0.3in}{0in}
    \phantomcaption{\label{fig:NAS10000}}
  \end{subfigure}
  
\vspace{-\baselineskip}
  
  \caption{\label{fig:KaikkiProd}The number of aftershocks per mainshock as a function of the mainshock's size and the threshold used. In (a) all events after a mainshock are counted as aftershocks, until there is an event at least the size of the mainshock. The continuous lines are fits using the function \ref{eqn:NAS}. Figures (b), (c), and (d) require the aftershock sequences to last for at least 100, 1000 and 10 000 timesteps respectively, and no further shocks are recorded. A power-law region becomes more apparent for larger time windows and thresholds. For large time windows and small thresholds there are no data for the small mainshocks, as none of their aftershock sequences are long enough for the time window. The continuous lines are fits with a function $\propto (S^\alpha-1)/(S^\alpha+S_{P}^\alpha)$, where $S$ is the mainshock's size and $S_{P}$ and $\alpha$ are constants. Again, the graphs showing data for different threshold have been moved vertically to avoid overlapping, and the legends show the thresholds that were used and the fitted exponents.}

\end{figure*}

The relationship between the size of a mainshock and the number of aftershocks turns out to be slightly complicated in our system. Doing a similar analysis as in the previous studies \cite{Bar_s_2018, Bar_s_2018b, Bar_s_2019} with the definition that all the events after a mainshock before another at least as large event are aftershocks, we find that the number of aftershocks grows as a power of the mainshock's size, as is shown in Figure \ref{fig:NAS}. 

This apparent productivity law does not describe how many shocks a mainshock triggers, but rather for how long does the defined aftershock sequence last. As larger avalanches are more scarce, there are of course more shocks between two large mainshocks than two small ones on average. Similarly, there should be more events between longer avalanches and more events between rare events in general. 

It is worth mentioning that the aftershocks in Figure \ref{fig:NAS} are not aftershocks in the same sense as in seismology, as their frequency does not necessarily follow the Omori-Utsu law for the duration of the whole sequence. Looking at Figure \ref{fig:Omori}, we see that the Omori-like aftershock sequences in our system last roughly for 10-10000 timesteps depending on the threshold.

As was shown in \cite{Bar_s_2018, Bar_s_2018b, Bar_s_2019}, with this definition of aftershocks the productivity law does not change even after randomly permuting the events. The authors found that the behaviour indeed follows from the ratio between the number of events smaller than a mainshock and the number of events at least as large as the mainshock. 

In Figure \ref{fig:NAS} the number of aftershocks for an avalanche of size $S$ is fitted using the integral of a size distribution of the form $(1+S/S_{min})^{-\tau_S}$ to get the number of events smaller than $S$ divided by the number of events with size $S$ or larger.
The resulting aftershock number 
\begin{equation}
    \label{eqn:NAS}
    N_{AS} = \dfrac { (1+(S-1)/S_{min})^{1-\tau_S} - (1+S_0/S_{min})^{1-\tau_S} } { (1+S_1/S_{min})^{1-\tau_S} - (1+S/S_{min})^{1-\tau_S} }, 
\end{equation}
where $S_0=1$ is the lower boundary and $S_1$ the upper boundary of the integral. The values of $\tau_S$ in Figure \ref{fig:NAS} are indeed very close to the values in the size distribution in Figure \ref{fig:sizes} despite neglecting the exponential cutoff of the size distribution.

A different and probably more interesting way to look at the number of aftershocks is to use a time window. In Figures \ref{fig:NAS100}, \ref{fig:NAS1000}, and \ref{fig:NAS10000} the aftershocks are still smaller than the initiating mainshock, but the sequences have to last for at least some specific duration, and aftershocks are counted only for that time. If the window is for example 5000 timesteps, sequences where there is an event larger than the mainshock after 4000 steps are ignored, and only the first 5000 steps of a sequence that lasts for 6000 steps are looked at.

When the aftershocks are counted only for a set time, the behaviour divides into three categories. For short time windows, the aftershock number consists mostly of the increased activity in the shock frequency distribution shown in Figure \ref{fig:Omori}. Consequently, in Figure \ref{fig:NAS100}, where the aftershocks are counted for 100 time steps, the aftershock number increases more for graphs with a higher threshold. The graphs are fitted with a monotonously increasing function $\propto (S^\alpha-1)/(S^\alpha+S_{P}^\alpha)$, where $S_P$ is the value of the shock size $S$ at which the aftershock number starts to plateau. Without a threshold, the data do not follow a similar function, but instead the aftershock number decreases after some value of the mainshock's size.

For slightly larger time windows such as in Figure \ref{fig:NAS1000}, where the window is 1000 time steps, the aftershock number includes more of the average activity in the simulations, and hence the behaviour becomes more similar for all thresholds. All graphs can be fitted with the function $\propto (S^\alpha-1)/(S^\alpha+S_{P}^\alpha)$, with the exponent $\alpha$ around one.

In Figure \ref{fig:NAS10000} the time window is 10 000 steps. With a small threshold there are no small mainshocks with long enough aftershock sequences, so the graphs start at large mainshocks. With large thresholds the aftershock number increases for almost the whole range of shock sizes, and the exponents are again close to one.

Combining the findings in Figures \ref{fig:NAS100}, \ref{fig:NAS1000}, and \ref{fig:NAS10000}, we can deduce that a large avalanche in interface depinning is most likely followed by a large number of smaller avalanches on a variety of time scales. Increasing the detection threshold extends the effect to a wider range of avalanche sizes. 

It is important to note that the results do not say that a small avalanche is followed by a small number of events. Large avalanches can still be preceded by small ones, so that the events that follow the large avalanches also follow the preceding small avalanches. But if we ignore small events that build up to larger ones, then the larger an avalanche is, the more events it is followed by, as long as a detection threshold is used.

\section{Discussion}

We simulated the depinning of a long range elastic interface using a cellular automaton model. Avalanches in the movement were defined using various thresholds to study their effect. As the driving force balanced around the depinning point, the interface moved intermittently and avalanches could also be defined without a threshold. 

A threshold divides avalanches into separate events whenever the velocity of the interface visits below the threshold \cite{Jani_evi__2016}. Consequently, we found that higher thresholds increased small and short avalanches and decreased large and long ones. Thresholds close to the average velocity changed the exponents of the size and duration distributions by about 10 percent compared to the pure signal with no threshold.

The seismic-like clustering of avalanches discussed in previous interface studies \cite{Bar_s_2018, Bar_s_2018b, Bar_s_2019} was investigated to see if a detection threshold would affect it. We found that the power-law distributed frequency of aftershocks depends on the use of a threshold. With no threshold, the shock frequency initially increases with time, as more waiting times between events end. With a threshold however, the aftershock frequency starts at a higher value and decreases as a power of time until meeting some background event rate. A higher threshold decreases the background activity and makes the power-law region longer.

The results applied also if the aftershocks could be larger than the mainshock they followed, so in general we found that a threshold causes avalanches in interface depinning to cluster in time with a power-law frequency. This clustering is probably a natural result of the power-law distributed waiting times caused by a threshold shown previously in \cite{Jani_evi__2016}.

We studied also the dependence of the number of aftershocks on the size of a mainshock. The aftershocks were looked at for different time scales. The number of aftershocks was proportional to a power of a mainshock's size as long as the timescale was long enough or a threshold was used. For small timescales and no threshold, the aftershock number did not grow monotonously with the mainshock's size, but rather decreased and plateaued after some value. A larger threshold and a larger time window led to longer and more apparent power laws, with exponents close to one. 

The fairly simple detection threshold as well as the method for classifying avalanches could be modified to study local events or a local threshold. First, lonely events that do not have enough activity around them inside some space-time window could be filtered out. We have already done initial tests using this type of a local threshold, and the results seem to mimic what was found here with the global threshold. The next step is to also classify the avalanches using a space-time window to separate simultaneous but spatially distant events.

\bibliography{tiejosto}

\end{document}